# Generalized coupling in the Kuramoto model

G. Filatrella

Laboratorio Regionale SuperMat INFM-CNR, Salerno and Dept. of Biological and Environmental Sciences, University of Sannio, Via Port'Arsa 11, I-82100, Benevento, Italy

N. F. Pedersen

Oersted-DTU, Section of Electric Power Engineering, The Technical University of Denmark, DK-2800 Lyngby, Denmark

K. Wiesenfeld

School of Physics, Georgia Institute of Technology, Atlanta, Georgia 30332

We propose a modification of the Kuramoto model to account for the effective change in the coupling constant among the oscillators, as suggested by some experiments on Josephson junction, laser arrays and mechanical systems, where the active elements are turned on one by one. The resulting model is analytically tractable and predicts that both first and second order phase transitions are possible, depending upon the value of a new parameter that tunes the coupling among the oscillators. Numerical simulations of the model are in accordance with the analytical estimates, and in qualitative agreement with the behavior of Josephson junctions coupled via a cavity.

Synchronization of coupled nonlinear oscillators appears almost ubiquitously in our world[1], to an extent that any listing will prove rather incomplete and unsatisfactory. The description of many nonlinear oscillators with different natural frequencies that must compromise to a common frequency requires a statistical treatment that is complicated by the nonlinear nature of the single elements. Most progress has been achieved within the framework of the theorist's playground for mutual entrainment, the Kuramoto Model[2][3]. However, the Kuramoto model has a stringent characteristic: it assumes global coupling, with each oscillator coupled to all others on an equal footing through the sum of a periodic function (a trigonometric function in the originally proposed model) and whose strength is tuned by a constant $K$:

$$\dot{\vartheta}_i = \omega_i + \frac{K}{N} \sum_{j=1}^{N} \sin(\vartheta_j - \vartheta_i - \alpha) \qquad (1)$$

The intuitive idea behind the model is that each oscillator if uncoupled would oscillate at the frequency $\omega_i$, but the resulting phase difference with the others produces a restoring "force" (proportional to the coupling $K$), thus favoring, as much as the distribution $g(\omega)$ of the natural frequencies $\omega_i$ allows, synchronous motion. The standard analysis of the Kuramoto model accurately predicts, when compared with simulations, the degree of synchronization as a function of the coupling $K$ and the width of the distribution of the $\omega_i$. Usually one gladly pays the price of this simplification (global coupling) to be compensated by the analytical insight offered by the model. But sometimes the situation is better -- for example when the cross talk between oscillators is mediated by a common connection to passive elements, say a linear resonator -- since it is possible that the global coupling approximation is not merely a crude truncation of the coupling to some averaged mean field, so that the analytical advantages are given back by paying only a low cost of simplification. For example, Josephson Junctions (JJ) coupled through a lumped resonator are described by the following equations (in normalized units):

$$\ddot{\varphi}_i + \frac{1}{\sqrt{\beta}}\dot{\varphi}_i + I_i \sin\varphi_i = b - \dot{q}$$
$$\ddot{q} + \frac{1}{Q}\dot{q} - \Omega^2 q = \frac{1}{\beta_L}\sum_{j=1}^{N}\dot{\varphi}_j$$
(2)

and in the limit of weak coupling to the resonator Eq.s (2) reduce to the Kuramoto model (1)[4,5].

This connection offers hope that the Kuramoto paradigm might be directly and quantitatively tested in real Josephson arrays with resonantor-coupling architecture, whose performance is known to depend on the degree of synchronization[6,7,8,9].

In this paper, we argue that an existing experiment -- suitably reinterpreted -- is very nearly a realization of the Kuramoto model. The key difference is the way in which the experimental system is tuned across the transition. In the theoretical model (1) the natural control parameter is the coupling coefficient $K$, with $N$ held fixed. In contrast, the experiments in Ref.[8] systematically increased the number of (active) oscillators. This difference leads us to propose a particular modification of the Kuramoto model which embodies the physical consequences of this sequential activation for the resonator-mediated coupling. Our modified model is analytically tractable, and leads to some interesting predictions. Among these is the possibility of a first order transition at the onset of synchronization, a feature reported in the above-cited experiment. (In contrast, Eq. (1) leads to a second order transition.) Other experiments where the number of elements can be systematically tuned, such as laser arrays[10] and a recent recreation of London's Millenium Bridge instability[11] [12] may also be described by our generalized model.

The main feature observed in experiments of underdamped Josephson junctions[8] [9] is the existence of a threshold for synchronization, with a sudden jump to a finite amplitude of emission, much above the incoherent emission from unsynchronized arrays. This signals a sudden increase in the degree of coherence, which is measured in the framework of the Kuramoto model by the order parameter $r$:

$$re^{i\psi} = \frac{1}{N}\sum_{j=1}^{N} e^{i\vartheta_j} \qquad (3)$$

that (roughly speaking) describes the fraction of locked oscillators (an incoherent state results in $r=0$, while a perfectly coherent state implies $r=1$). The Kuramoto analysis predicts that increasing the strength of the coupling from zero, has no effect on the order parameter up to a critical value of the interaction $K_c=2/[\pi g(0)]$, and above this value $r$ increases monotonically toward the perfect synchronous state $r=1$, so that it is natural to describe the transition as a *second order* phase transition. However, simulations of system (2), adding the active oscillators one by one, reveal a different behavior of the order parameter $r$ (see Fig. 1). The sudden jump in $r$ at onset is just the feature seen in the experiments[8]. A few years ago [13], we proposed that this difference could be attributed to three reasons: a) The oscillators are underdamped, and Kuramoto models with inertia lead to a first order phase transition[14]; b) the center of the distribution of the natural frequencies might have been displaced respect to the center of the cavity ($\Omega=<\omega_i>-\delta$); and c) the oscillators are activated one by one. However, it has since been shown that underdamped Josephson junctions are also described by a massless Kuramoto model[5]; moreover, our present simulations of Josephson arrays with frequencies centered around the peak of the resonance, $\delta=0$, as those of Fig. 1, demonstrate that item b) is not a fundamental requirement, although it might play a role in the quantitative analysis. We therefore conclude that the addition of oscillators one by one is the main cause of the sudden jump to a finite value of the order parameter $r$. Since the physical consequence of adding active junctions is to increase the amplitude of the resonator oscillations (see Eq.(2)), we therefore are led to variation of the Kuramoto model that accounts for the change of the coupling with the number of active oscillators, i.e.:

$$\dot{\vartheta}_i = \omega_i + \frac{K(r)}{N}\sum_{j=1}^{N}\sin(\vartheta_j - \vartheta_i) \qquad (4)$$

A simple functional form for $K(r)$ to describe this effect is the power-law $K(r)=Kr^{z-1}$. In this general framework the Kuramoto model is the special case of constant coupling $z=1$, for $z>1$ one gets models in which the coupling is enforced when more rotators are synchronized, while for $z<1$ the coupling is weakened. The actual value of $z$ will depend on the parameters of the model (2) in some complicated way, because the tendency of the active oscillators to attract to each other will vary with the parameters of the model. We will therefore use the parameter $z$ as a measure of the strength of the feedback mechanism, which might also account for the shift $\delta$ of the natural frequencies $<\omega_i>$ with respect to the peak $\Omega$ of the resonator. Simulations of the Eqs (4) for various values of the exponent $z$ are reported in Fig. 2, together with the analytic predictions to be derived below. Several features are clear from the figures:

1. For $z \leq 1$ the evolution from the incoherent value $r=0$ to the partially coherent state is continuous. Strictly speaking, there is no transition except for the special case $z=1$ where the expected second order phase transition is retrieved; for $z<1$ the order parameter is small but finite, and reaches zero only for $K=0$.

2. The finite accuracy of the numerical simulations masks this difference between the special case $z=1$ and the $z<1$ cases. Presumably, in real systems a smooth, second order-like transition would be observed also for systems best described by a negative feedback $z<1$.

3. For $z>1$, as the parameter $K$ is increased past some critical value $K_c$, there is an onset of synchronization accompanied by a jump to a finite value of the order parameter ( ''first order phase transition'' ).

4. For $z>1$, as the parameter $K$ is decreased the numerical simulations show hysteretic behavior, with the transition back down to the $r=0$ state occuring at a lower coupling than $K_c$.

We note that our system for $z>1$ behaves qualitatively similar to more traditional non-linear systems where the analogy between bifurcations and phase transitions can be drawn. A well known and simpler system that behaves qualitatively similar is the parametric excitation of half harmonic oscillations in the rf-driven Josephson junction [15] [16]. Here there is only a single non-linear element, and the strength of rf-amplitude takes the role of the strength of the coupling constant (i.e. the X-axis). The half harmonic amplitude takes the role of the ''order parameter'' (Y-axis) while the parameter $\omega/2\omega_p$ takes the role of the parameter a determining the qualitative behavior of the phase transition. The period doubling bifurcation in a one dimensional iterative map[16] provides a similar example. We also note that the effect of noise close to the onset of the phase transition in the coupled oscillator system is qualitatively similar to the behavior of the single driven Josephson junction system. In the latter case there appears the phenomenon of noise rise near the onset of the phase transition that many authors have studied experimentally.

To analytically handle Eq.s (4) we cast them in the form:

$$\dot{\vartheta}_i = \omega_i + K(r)\sin(\psi - \vartheta_i) = \omega_i + Kr^z \sin(\psi - \vartheta_i). \tag{5}$$

(For $z=1$ this is just the standard Kuramoto model, for $z=2$ see[17].) Following the usual analysis[18], let us suppose that there exists a solution where $r$ and $\psi$ are time-independent, and that the synchronous state rotates at the peak frequency of the distribution $g$, i.e. in our reference system it is still ($<\omega_t>=0$). From Eq. (5) we get that the synchronized oscillators are just frozen with a fixed phase

$$\omega_i = Kr^z \sin(\vartheta^*_i) \Rightarrow \vartheta^*_i = \sin^{-1}(\omega_i / Kr^z) \tag{6}$$

Put another way, each oscillator is accommodated in a phase that depends upon the natural frequency of that specific oscillator ($\omega_i$) and the global property of the synchronized state, the actual and unknown order parameter $r$. Another relevant consequence of Eq. (6) is that, depending on $K$, there will be a maximum frequency $\omega$ that can be synchronized. So one can divide the distribution $g(\omega)$ in two parts, a portion around zero of width $Kr^z$ that participates in the synchronous motion (in this reference frame these phases are constant) and the outliers that rotate. Following Kuramoto we assume that an even $g(\omega)$ guarantees that the order parameter (2) evaluated for the outliers is zero (as many rotate clockwise as rotate counterclockwise), so the only nonvanishing contributions come from the central portion of the oscillators. To estimate $r$ for sufficiently many oscillators one estimates (2) with the integral:

$$re^{i\psi} = \frac{1}{N}\sum_{j=1}^{N} e^{i\vartheta_j} \cong \int_{-K'r^z}^{K'r^z} e^{i\vartheta} g(\omega)d\omega = Kr^z \int_{-\pi/2}^{\pi/2} e^{i\vartheta} g(Kr^z \sin\vartheta)\cos\vartheta d\vartheta =$$
$$= Kr^z \int_{-\pi/2}^{\pi/2} \left[g(Kr^z \sin\vartheta)\cos^2\vartheta + ig(Kr^z \sin\vartheta)\cos\vartheta\sin\vartheta\right]d\vartheta \quad (7)$$

The second part of the integral is zero if $g$ is even, so the self-consistency condition for the drifting oscillators requires that either $r = 0$, or

$$1 = K'r^{z-1}\int_{-\pi/2}^{-\pi/2}\cos^2(\vartheta)g(K'r^z \sin\vartheta)d\vartheta \quad (8)$$

Assuming a Lorentzian distribution $g(\omega)=(\gamma/\pi)/(\gamma^2+\omega^2)$ the integral condition (8) is readily evaluated to give:

$$K = \frac{2\gamma r^{z-1}}{(1-r^2)}. \quad (9)$$

The critical value of the coupling $K_c$ can be retrieved observing that it corresponds to the minimum coupling, $K_c$: $\partial K/\partial r = 0$, so differentiating (9) one gets:

$$r_c = \frac{\sqrt{z-1}}{\sqrt{z+1}} \quad (10a)$$

$$K_c = \frac{2\gamma\left(\frac{\sqrt{z-1}}{\sqrt{z+1}}\right)^{z-1}}{1-\frac{z-1}{z+1}} \quad (10b)$$

Equations (9) and (10) correspond to the branch of coherent states ($r > 0$). The predictions (10) are compared with numerical simulations in Fig.3. The estimates depend, for a given set of random frequencies extracted from a Lorentzian distribution, upon the specific order in which they are activated, and therefore result in very large fluctuations.

The comparison with numerical experiments such as those of Fig. 1 reveals that, at least for some values of the coupling with the resonator, the transition to the synchronous state is very similar in the Josephson junction array and in the Kuramoto model with $z>1$.

In conclusion, we have demonstrated that if globally coupled Josephson junctions are activated one by one, a description in terms of the usual mapping onto the Kuramoto model is inadequate, while a modified model in which the coupling depends (via a new parameter $z$) on the fraction of synchronized junctions is more appropriate. Physically, the parameter $z$ controls the degree of feedback provided by the coupling resonator. The modified Kuramoto model remains analytically tractable, and our analysis shows that the standard case $z=1$ is rather special. For $z<1$, which corresponds to a weakening of the coupling by increasing the order parameter $r$, no phase transition is predicted although the order parameter stays at a vanishingly small value. For $z>1$, which corresponds to a re-enforcing of the coupling by increasing the order parameter $r$, a first order phase transition is predicted.

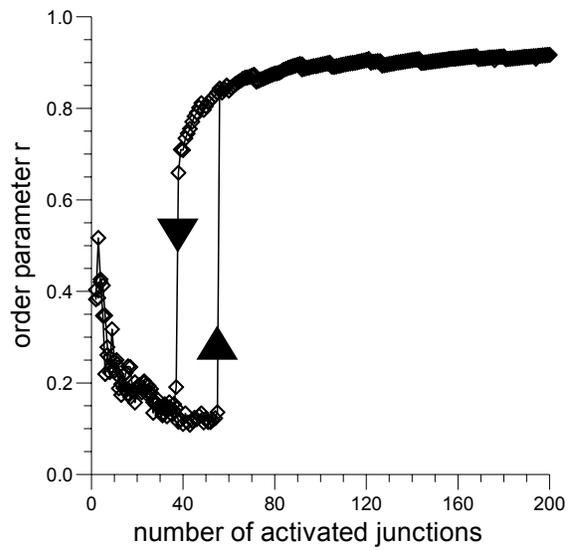

**Fig.1: Behavior of the order parameter for a Josephson Junction array coupled to a cavity obtained by increasing and decreasing the number of active junctions. Parameters of the simulations are: *N=200, b=0.6, β=10, β$_L$=5000, Q=100, Ω=<ω$_i$>*. The disorder is set to *γ=0.05*.**

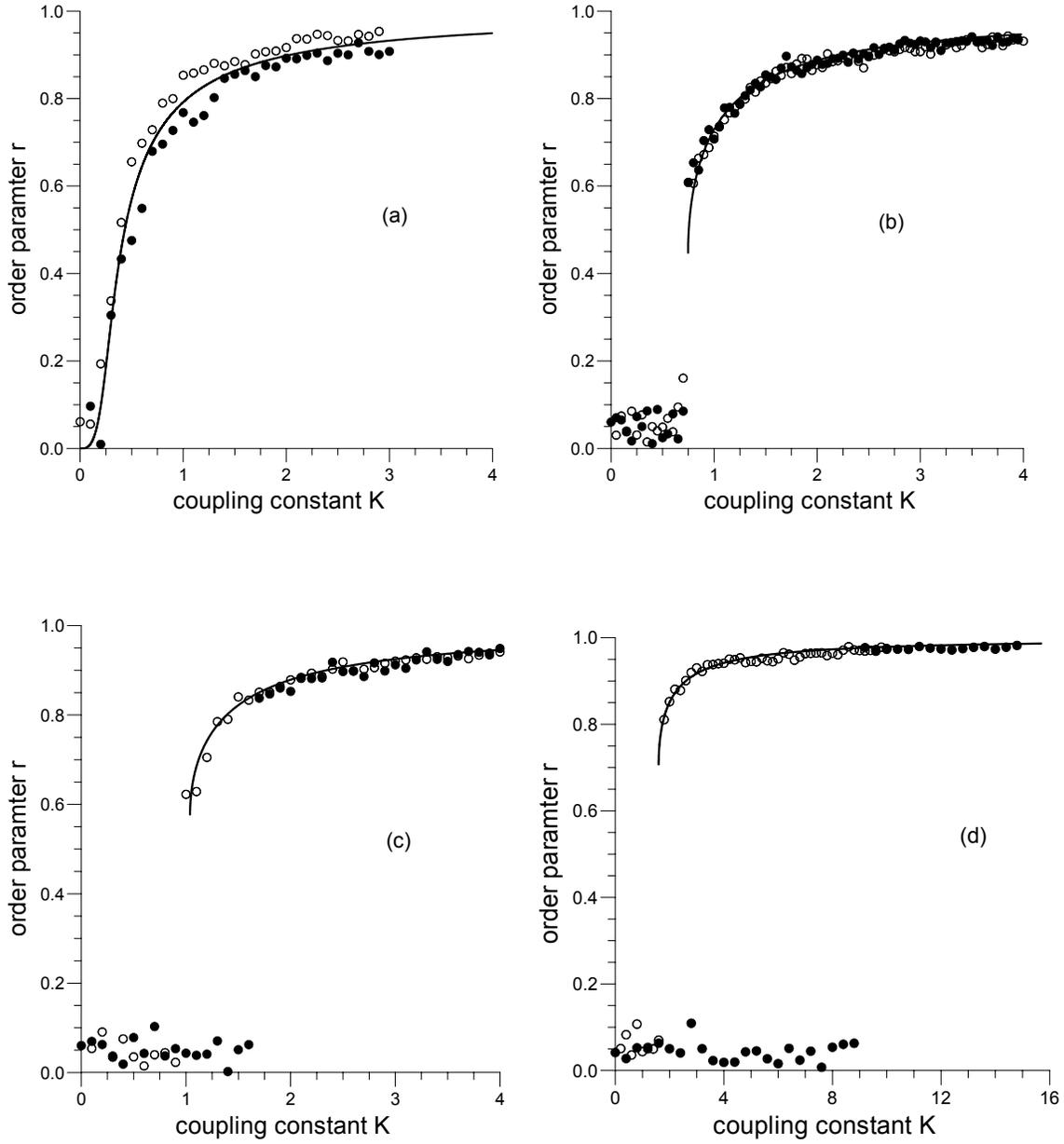

**Fig. 2:** Numerical and analytical results of Eq. (3) for various feedback strengths: $z=0.7$ (a), $z=1.5$ (b), $z=2$ (c), $z=3$ (d). The solid lines represent the analytic prediction, the symbols represent numerical data obtained by increasing (open circles) and decreasing (filled circles) the coupling. The disorder is set to $\gamma=0.05$.

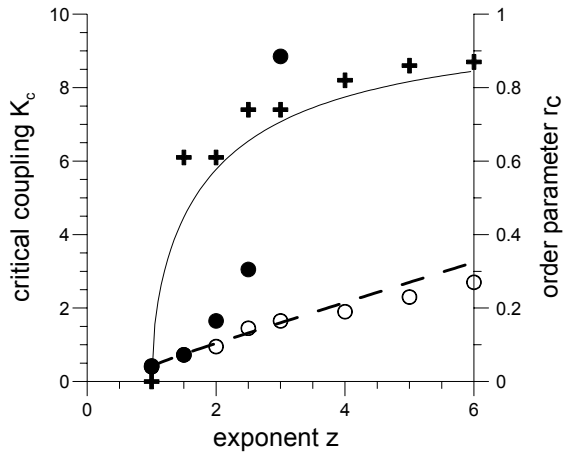

**Fig. 3: Minimum value for the coupling parameter *K_c* (dashed line, left axis) for the onset of synchronization and the corresponding value of the order parameter *r_c* (solid line, right axis). Circles refer to the numerically determined critical coupling obtained by increasing (filled symbols) and decreasing (open circles) the coupling. Crosses refer to an estimate of the order parameter at the onset of synchronization. The disorder is set to *γ=0.05*.**

---

[1] S. Strogatz, *Synch: The emerging Science of spontaneous Order* (Hyperion, New York, 2003).

[2] Y. Kuramoto, in *International Symposium on Mathematical Problems in Theoretical Physics*, edited by H. Araki, Lecture notes in Physics No. 30 (Springer, New York, 1975).

[3] J.A. Acebrón, l.L. Bonilla, C.J. Perez Vicente, F. Ritrot, and R. Spigler, Rev. Mod. Phys. **77**, 137 (2005).